\documentclass[10pt,letterpaper]{article}
\usepackage[top=0.85in,left=2.75in,footskip=0.75in]{geometry}

\usepackage{amsmath,amssymb}

\usepackage{changepage}

\usepackage{textcomp,marvosym}

\usepackage{cite}

\usepackage{nameref,hyperref}

\usepackage[right]{lineno}

\usepackage[nopatch=eqnum]{microtype}
\DisableLigatures[f]{encoding = *, family = * }

\usepackage[table]{xcolor}

\usepackage{array}

\newcolumntype{+}{!{\vrule width 2pt}}

\newlength\savedwidth

\newcommand\thickhline{\noalign{\global\savedwidth\arrayrulewidth\global\arrayrulewidth 2pt}%
\hline
\noalign{\global\arrayrulewidth\savedwidth}}

\raggedright
\usepackage[aboveskip=1pt,labelfont=bf,labelsep=period,justification=raggedright,singlelinecheck=off]{caption}

\makeatletter
\renewcommand{\@biblabel}[1]{\quad#1.}
\makeatother

\usepackage{lastpage,fancyhdr,graphicx}
\usepackage{epstopdf}
\fancyheadoffset[L]{2.25in}
\fancyfootoffset[L]{2.25in}
\begin{document}
\vspace*{0.2in}


\begin{flushleft}
{\Large
\textbf\newline{Quantitative analysis of the value of investment in research facilities, with examples from cyberinfrastructure} 
}
\newline
\\
Winona G. Snapp-Childs\textsuperscript{1},
David Y. Hancock\textsuperscript{2},
Preston M. Smith\textsuperscript{3},
John Towns\textsuperscript{4},
Craig A. Stewart\textsuperscript{5*}
\\
\bigskip
\textbf{1} Pervasive Technology Institute, Office of the VP for IT, Indiana University, Bloomington, IN, USA
\\
\textbf{2} Research Technologies, Office of the VP for IT, Indiana University, Bloomington, IN, USA
\\
\textbf{3} Rosen Center for Advanced Computing, Purdue University, West Lafayette, IN, USA
\\
\textbf{4} National Center for Supercomputing Applications, University of Illinois Urbana-Champaign, Urbana, IL, USA
\\
\textbf{5} Department of Computer Science, Indiana University, Bloomington, IN, USA

\bigskip

%
%





* stewart@iu.edu or cas@pobox.com

\end{flushleft}
\section*{Abstract}
Purpose: How much to invest in research facilities has long been a question in higher education and research policy. We present established and recently developed techniques for assessing the quantitative value created or received as a result of investments in research facilities. This discussion is timely. Financial challenges in higher education may soon force difficult decisions regarding investment in research facilities at some institutions. Clear quantitative analysis will be necessary for such strategic decision-making. Further, institutions of higher education in the USA are currently being called on to justify their value to society. The analyses presented here are extendable to research enterprises as a whole. 

Results: We present methods developed primarily for analyses of cyberinfrastructure. Most analyses comparing investment in university-based cyberinfrastructure facilities with purchasing services from commercial sources demonstrate positive results for economic and scientific research. For example, one recent analysis showed a positive relationship between investment in cyberinfrastructure facilities and high-impact publications and number of PhDs conferred by a university. Another assessment showed that a publicly funded cyberinfrastructure project delivered to the USA economy and society exceeded the cost to USA taxpayers. 

Conclusions: Quantitative analyses of the benefits of investment in research and research facilities create a fact-based foundation for discussing the value of research and higher education. These methods enable a quantitative assessment of the relationship between investment in specific research facilities or research projects and economic, societal, and educational outcomes. These methods are of value in quantifying the economic benefit of higher education and in managing investments within such institutions. They may be applied retroactively, making this report of potentially immediate value for institutional assessments. Use of methods for depicting value developed for Total Quality Management may be valuable to depict such assessments.



\section*{Introduction}
Institutions of higher education are called on to educate, to engage in research, to create (scientific breakthroughs, works of art), and to contribute to the public good. Every college and university, whether public or private, has its own mission and priorities attached to these core activities. Since the beginning of the modern era of research activities, which in the USA began with the 1945 report by Vannevar Bush entitled “Science, the Endless Frontier" \cite{Bush1945}, there have been very few times during which some universities were able to spend money at will. During most of the past 80 years, availability and allocation of funding has been a matter of some concern to almost all institutions of higher education. This paper presents longstanding and new methods for the evaluation of investment in research-enabling facilities, specifically focusing on methods developed for analysis of computing infrastructure used to enable research activities. All of these methods are in principle extendable to investment in research generally.

This topic is particularly timely in the USA. There have been for many years projections of crises in higher education budgets due to changes in demographics and, in the case of public universities, levels of funding from their home states. These issues have become critical for many institutions of higher education in the past few years. Colleges and universities in the USA closed at a rate of more than one per month in 2024 \cite{Moody2024}. In 2023 and 2024, at least 10 universities in the USA - more than one per quarter - announced a budget deficit exceeding USD 10M (see S1 Table 1 for the full list of these institutions). Universities announcing large budget deficits included R1 universities such as the University of Nebraska, West Virginia University, and the elite Ivy League school Brown University. (R1 - research 1 -  universities are described as “doctoral universities” with “very high research activity” in the Carnegie classification of higher education institutions in the USA \cite{Carnegie2023}). While presently less critical, budget concerns are also present in European higher education institutions \cite{EuropeanCommission2024}. 

In 2025 a new sort of financial challenge has emerged. The USA federal government has threatened dozens of institutions of higher education with the loss of grant funds due to alleged violations of the USA Civil Rights Act of 1964. Some universities are at risk of having hundreds of millions to billions of dollars in total grant funding canceled \cite{Waldenberg2025}. Nothing in this paper will be relevant to the resolution of actual legal misdeeds. However, if, as some have asserted, the lodging of these allegations rests on an underlying ideological questioning of the value of higher education in the USA, then demonstration of benefits to the USA as a whole that derive from research conducted at institutions of higher education may be of use in discussions. This is not just a problem in the USA. Although this distinct legal issue may be specific to the USA, ideological questions about the value of higher education are not.

When financial challenges become an existential threat to institutions of higher education, every expenditure must be analyzed and potentially subject to drastic change. Even when institutions of higher education have weathered financial crises, such challenges can have significant impact on university structure and mission. For example, West Virginia University eliminated 143 faculty positions and 28 academic programs as a result of its recent budget shortfall\cite{Quinn2024}. Such decisions are grave and must be made based on the best possible information. 

The methods highlighted in this paper for quantifying the relationship between investment in research facilities and financial, societal, and educational outcomes can be valuable in setting research policy at institutions of higher education in normal planning times, as well as in times of financial stress such as many institutions of higher education are currently facing. The older methods we present are rooted in the Total Quality Management movement that began in the 1980s and the somewhat newer but time-tested discipline of bibliometrics.

More recent methods have been developed specifically for cyberinfrastructure.  In the USA, the term cyberinfrastructure is used for research computing facilities and services defined as encompassing “computing systems, data storage systems, advanced instruments and data repositories, visualization environments, and people, all linked together by software and high performance networks to improve research productivity and enable breakthroughs not otherwise possible” \cite{Whatiscyberinfrastructure2010}. Cyberinfrastructure is thought by many to be of use in supporting research at higher education institutions. However, one vice-president for research, in stating his general belief in the value of cyberinfrastructure, noted that this was an article of faith for him, not something he was able to prove empirically. And of course, higher education cyberinfrastructure facilities are not invulnerable to budget pressures. Recent cuts in university cyberinfrastructure facilities and staff are documented in \cite{Snapp-Childs2024}. The value of investment in cyberinfrastructure has been a topic of considerable investigation because there have been options both in terms of how much to spend and how to spend. Since 2006, institutions have had the option of purchasing and operating cyberinfrastructure facilities or outsourcing delivery of such facilities to a commercial cloud provider \cite{AWS2025}. While it is hard, for example, for an institution of higher education to outsource something like a high-powered MRI facility or laboratory animal facilities, the increasing ease with which an institution of higher education can outsource cyberinfrastructure has inspired particular focus on the quantitative analysis of optimal investment mechanisms for computing facilities. 

It would make little sense to disseminate information about methods for analyzing the value of investing in cyberinfrastructure if the results had been unclear or equivocal - but this has not been the case. Research on the cost-effectiveness of on-premises cyberinfrastructure systems shows that outsourcing computational activities would have been three times more expensive for a university than operating an on-premises supercomputer \cite{Stewartetal2018}. In terms of investments supporting national research communities, it would have been two to three times more expensive for the USA federal government to contract with commercial firms than to contract with universities to provide cloud services \cite{Stewartetal2018} or cyberinfrastructure support and management services for a national network of advanced computing resources \cite{Stewartetal2023}. Additional research shows strong relationships between investment in local computing facilities and support staff, on the one hand, and numbers of new PhDs and total university research spending (closely related to total grant income to the university), on the other \cite{Smith2024} \cite{Smithetal2025}. 

These recently developed methods for quantifying the value of investment in cyberinfrastructure are straightforwardly extensible to any type of research-enabling facility investment. It may also be possible for some of these methods to extend the analysis of investment to research enterprises more generally and to be applied after the fact, so long as good organizational records exist. Given this broader reach, these methods should be of value to researchers, research administrators, and senior institutional leadership.

National policies related to research at institutions of higher education are tied to success in the commercialization of new technology products that improve quality of life, global competitiveness and defense, and profits \cite{Mandt2020} \cite{NASEM2010}. Leaders at institutions of higher education are at the vanguard of a technology ecosystem that may take decades to turn an idea into a widely used product. Because of the time lag between a new discovery and its translation into improving lives, stakeholders of all sorts need good information to enable ongoing evaluation during this development and translation process. It is also important to note that while the methods presented here arise from the current challenges being faced in higher education in highly developed countries, they are equally relevant in justifying expansion of research activities in academia in the Global South. 

In the remainder of this paper, we summarize recent developments in the quantitative assessment of the value of research facilities, with an emphasis on methods recently developed in studies of cyberinfrastructure. We briefly consider implications for the recent results of studies focused on the value of cyberinfrastructure. Although we do not present any new data, reconsidering the result of five recent publications of the authors together yields new insights. We then discuss the extension of the methods to other kinds of research-enabling facilities and research initiatives generally. The methods presented here were developed for evaluation of cyberinfrastructure in a higher education setting, but are applicable to research and development in any setting. We show an example of how the now decades-old Balanced Scorecard concept \cite{Kaplan1992} may be adapted to concisely convey quantitative metrics relevant to research-enabling facilities and research generally. We then conclude with a few comments about potential areas of future research.  

Different kinds of readers are likely to draw different insights from this paper. We hope that leaders and staff of research-enabling facilities and researchers in general will see new ways to express the value of their activities. We hope that a skeptical CFO (Chief Financial Officer) might take this paper to a colleague in economics or accounting and receive back the assessment, “within the bounds of the data that are available, the methodology and conclusions are reasonable.” We hope that any senior leader in higher education will find, for purposes of communicating outside their home institution, new means to explain the value of their institutions to important stakeholders and, for purposes within their home institution, helpful approaches for financial and strategic decision making. We dearly hope that researchers and higher education administrators in the Global South will find tools that are applicable in their contexts to aid their work. Last, but not least, we hope that researchers in the areas of economics of innovation and science of science will find here sources for fresh ideas in these important areas.

\section*{Prior Related Research}
Economic studies that attempt to draw a general link between research and economic benefit go back more than a century, to the 1912 publication of Schumpeter's \textit{Theorie der wirtschaftlichen Entwicklung (Theory of Economic Development)} \cite{Schumpeter2003}. Qualitative analyses have long pointed out relationships between investment in research and economic growth. A current example of this sort of approach, used in the USA since at least the 1990s, can be found in \cite{NASEM2020}. This sort of analysis links broad areas of economic activity, such as telecommunications, back to specific suites of USA federal government investment in research. While these qualitative relationships are convincing, they are drawn out only at very large scales and only in rare cases point back to a single research-enabling facility as a causal factor. As a result, such approaches are not generally helpful in evaluating an individual research-enabling facility and not at all helpful in managing budgets within institutions of higher education.

More recent research relevant to the economic benefits of research generally and investment in research facilities specifically have been examined more quantitatively. Three papers and one book provide particularly important and pathbreaking analyses of this relationship.  

In 2001, Salter and Martin \cite{Salter&Martin2001} reviewed the economic benefits of publicly funded basic research. Salter and Martin describe several different conceptual views of basic research. Prominent in their discussion was consideration of information-based economic theories. In this approach, one analyzes observed economic growth, associates as much as possible with quantifiable economic factors, and the difference is attributed to new discoveries and information. Salter and Martin point out an important distinction between private returns, namely “return on investments in research that flow ... to the organization directly involved,” and social returns, which are “the benefits which accrue to the whole society.” 

In 2016, Del Bo reviewed the rate of return of large-scale “research infrastructures,” which is a term from the literature in the European Union \cite{Del-Bo2016}. The term “research infrastructure" is derived from Florio's early work and is defined as “high-capital intensity and long-lasting facilities and equipment, typically operating in oligopoly conditions, whose objective is to support economic development and produce social benefits through the generation of new knowledge and, often, other spillover effects” \cite{Florio&Sitori2014}. Del Bo's analysis of return on investment (ROI) in research infrastructures helpfully creates a categorization of studies based on scope, ranging from macroeconomic to project-level studies \cite{Del-Bo2016}. Del Bo's excellent review of macroeconomic methods applicable to research infrastructures is based largely on endogenous growth models. Del Bo's efforts were focused on the European Union, but the methods outlined are broadly applicable to any situation involving an appropriately large and long-lasting research infrastructure. 

The European facilities of CERN (\textit{Conseil européen pour la recherche nucléaire} - the Europe-based organization for Nuclear Research) - is perhaps the best-known example of a persistent and large-scale Research Infrastructure in the sense of Del Bo's definition. The investments in CERN are so large that it is possible to assess their impacts through national and multinational macroeconomic studies. Florio followed his early paper-length analyses with an extensive book devoted to “social cost-benefit of research infrastructures" \cite{Florio2019}. This book offers a thorough conceptual overview of types of benefit that may derive from investment in large-scale Research Infrastructures and quantitative methods for estimating such effects. In summary, research into the public value or public investment in CERN shows it has a net positive impact on society. Many other of the Research Infrastructures reviewed by Del Bo also have a net positive economic impact.

The first peer-reviewed publication relating investment in advanced computing resources to a university's economic and academic success was published in 2010 \cite{Apon2010}. Apon's independent variable was investment by an institution of higher education in a high performance computing (HPC) system of computational capability sufficient to be included in the list of the 500 fastest supercomputers in the world \cite{Strohmaier2024}. Apon’s analysis included data from the inception of the Top 500 list in 1993 to 2009 during which time the number of higher education institutions in the USA on the list rose from 49 to 100. This analysis demonstrated that institutional investment in cyberinfrastructure was associated with increased peer-reviewed publications and grant income. This was a breakthrough in the quantitative analysis of investment in cyberinfrastructure related to research outcomes. Unfortunately, the sort of analysis done by Apon is not as helpful today as it was when her 2010 paper first appeared. The most recent Top 500 list, published in November 2024, includes just 15 computing systems funded and owned by an institution of higher education in the USA. It has become extremely expensive to purchase a system with sufficient capability to make it onto the list. The acquisition cost of the system ranked 500th on the most recent list was approximately USD 5M. The 15 systems funded and owned by universities on the list represent very few data points for analysis, and one would have to extrapolate to systems smaller than those on the list in order to draw conclusions relevant to most institutions of higher education. Perhaps most importantly, while Apon's analysis shows that increased investment in HPC systems led to increased grant income and increased output of peer-reviewed publications, it did not establish correlation coefficients that would answer the question, “Is investing in HPC facilities a net financial positive for universities?” 

Still, Apon's initial work and the 2014 panel session led to the first publication to offer a quantitative analysis that asserted the return on investment in a cyberinfrastructure project was greater than 1.0 \cite{Stewartetal2015}. That 2015 paper was the first iteration of ongoing analyses in the last decade that led to the current report.

\section*{Scope, Methods, and Critical Results}
CERN and cyberinfrastructure facilities at universities in the USA have been the subject of more cost-benefit and ROI analyses than most - perhaps any - research-enabling facilities in the world. Most cyberinfrastructure facilities involve orders of magnitude less investment and less economic impact. As a result, their impact is generally not detectable via the macroeconomic approaches used in analysis of CERN and other large-scale Research Infrastructures. Recognizing the extensive use of the term Research Infrastructure as defined by Florio and implying persistent and very large-scale facilities, we will focus on activities at a smaller scale. We use the term research-enabling facility, as the name implies, to signify facilities at scales smaller than a Research Infrastructure as defined by Florio.


In this section, we present project-level approaches to quantitative assessment of investment in cyberinfrastructure. Much of the research cited in this current paper is project-level analysis of one of the largest cyberinfrastructure projects ever funded by the USA's National Science Foundation (NSF), called XSEDE - the eXtreme Science and Engineering Discovery Environment. XSEDE included supercomputers, visualization resources, data storage facilities, and training and support services, described in detail in \cite{Towns2014} and \cite{Stewartetal2023}. XSEDE operated from 2011 to 2022, providing services supporting non-classified research by the USA national research community. Because of its scale and visibility, the value provided to the USA by XSEDE has been examined with notable care. 

Our discussion of project-level approaches to evaluating research-enabling facilities uses a categorization based on that presented by Del Bo and expanded to include other and newer analytic methods:

\begin{itemize}
    \item quality management approaches;
    \item bibliometric analyses;
    \item time savings;
    \item econometrics-based methods for estimating financial impacts and outcomes;
    \item accounting-based methods for estimating financial impacts and outcomes:
    \begin{itemize}
        \item leverage;
        \item a proxy for ROI;
        \item and a comprehensive framework for assessing net financial value creation;
    \end{itemize}
   
    \item and a production-function analysis of academic and financial outputs.
\end{itemize}

Note that in every area save two, our discussion includes research results by one or more of the authors on this paper except: discussion of time savings, which we think is interesting and bears mention even though we have not done this sort of work ourselves; and bibliometric analyses, an area in which work has been done under subcontract from coauthor Towns.

We focus on the critical information that potentially skeptical stakeholders, including government officials and the general public, commonly ask for: Is the work enabled by a research-enabling facility sufficiently \textit{financially} significant that the investment in that facility is \textit{financially} worthwhile and beneficial to society as a whole? 

\subsection*{Quality management approaches}
The total quality management movement reached the USA with force in the 1980s. Total quality management is “a term first used to describe a management approach to quality improvement” \cite{ASQ2025} and is explained in detail in \cite{George1992}. One focus of total quality management is the creation and measurement of a set of metrics that meaningfully define quality and costs for an organization. Such metrics have long included client or user reported satisfaction, total cost of ownership (TCO), and usage metrics that are typically specific to organizational type and an organization's specific goals. Such approaches have been applied to computing services in academia since at least the 1990s  \cite{Peeblesetal2001}.  All of these metrics are useful, but they have limitations. 

Client satisfaction surveys are a basic aspect of the quality assessment of any research-enabling infrastructure or facility. (Term usage varies. “Client” is regarded as a respectful term among many who are steeped in TQM methodologies, although the term “user” is also commonly used.) Of course, highly negative survey ratings indicate problems. Positive ratings demonstrate that the users of a facility like it, which is good. However, positive satisfaction scores do not necessarily prove that a facility is a cost-effective or strategic investment \cite{Peeblesetal2001}. 

The total cost of ownership (TCO) is a critical cost metric. TCO estimates for operating a research-enabling facility must include the operational costs of staffing, electricity, cooling, maintenance, and insurance, along with the prorated or amortized costs of capital purchases. While it is important to understand the TCO, in isolation from quality metrics it is just a budgeting figure that provides little information about the value derived from the cost.

For many categories of research-enabling facility there are also specific sorts of traditional operational metrics. A specific organization may also have certain goals tied to strategic objectives. Typical metrics for the operation of cyberinfrastructure facilities include percentage uptime (time during which a system was available for use as a percentage of theoretical maximum), the number of jobs submitted and completed, CPU- and GPU- hours used per unit, and the amount of data stored. Again, all of these metrics must be carefully interpreted. Unusually low usage numbers relative to resource availability may be indicative of some sort of problem - perhaps there has been insufficient training, or the system is not perceived by users to match their needs. Low usage statistics may also be found immediately after a large upgrade to the capability of computational systems when the people who use the system have not adjusted to the increased availability of cycles. Consistent usage near peak capacity may be indicative of a balanced match of availability and need. Consistent usage at peak capacity may indicate inadequate supply, or a doctoral student iteratively reanalyzing their data to avoid actually writing their dissertation, or a faculty member running mindless parameter sweeps rather than running carefully thought out and efficient computational experiments. (These are not hypothetical examples; the last named author was responsible for HPC systems for 30 years and experienced every one of these examples at least once.)

There will be discipline-specific operational metrics appropriate to any sort of research-enabling facility. While such metrics can clarify the activities and utilization of a research-enabling facility among practitioners in a given area, these operational metrics are not easily translatable into terms that relate value to investment. As such, they are unlikely to resonate intuitively with certain important categories of stakeholders, potentially ranging from taxpaying voters to elected officials to University Presidents and CFOs (Chief Financial Officers).

\subsection*{Bibliometric analyses}
Bibliometric methods have long been used to analyze the impact of scientific publications. As stated in \cite{Vavryçuk2018}, the “simplest way ... to measure the quality of scientists is to evaluate the following three integer numbers: the number of published papers, the number of citations, and the h-index ... defined as the maximum number of papers of a scientist which are cited at least h times.” At the level of government research policy, Corredoira et al. used bibliometric analyses to show that within the USA, “breakthrough inventions” were primarily a result of research “conducted outside the federal government and sponsored by the DOD [Department of Defense], HHS [Department of Health and Human Services] and NSF” \cite{Corredoira2018}.

In terms of the analysis of the impact of research-enabling facilities, Del Bo demonstrates a clear, careful, thorough, and well-justified approach to calculating an intellectual return function for investment in research facilities, based on number of publications, reputation of the journals in which publications appear, and citations to those publications \cite{Del-Bo2016}. 

One example of analysis centered on research facilities is presented in von Laszewski et al. \cite{vonLaszewskietal2021}, who performed a bibliometric analysis of peer-reviewed scientific publications created with analyses of simulations performed on XSEDE cyberinfrastructure facilities. They investigated the question: Are research publications created using XSEDE cited more often than similar papers created without the use of XSEDE? Their analysis suggested that the answer was yes. Unfortunately, however, this analysis suffered from a significant conceptual shortcoming. To investigate the impact of a specific research-enabling facility on overall research output, one really ought to ask two questions: 1) Are papers produced using a certain type of research-enabling facility more impactful, in general, than similar papers produced without this type of research-enabling facility? 2) Among papers produced using one type of research-enabling facility, are those produced using one specific facility (or perhaps one set of instruments) more impactful than other papers produced using facilities of the same type (or instruments of the same category)? The line of analysis suggested by von Laszewski et al. is interesting, but a more careful analysis is needed, hopefully published in one of the standard journals that focus on bibliometrics. 

Useful as they may be within academia, bibliometric measurements - even if expressed in the format of a benefit function - don't easily translate into a financial value function and are, as such, unlikely to be clearly and intuitively understood as a tangible metric of the value of investment in research. 

\subsection*{Time savings}
Wiedmann measured the value of a research support facility in terms of researcher hours saved \cite{Wiedmann2017}. Wiedmann's initial analysis concerned a virtual lab that included several cyberinfrastructure resources. This approach could be very useful in determining staffing levels for a research-enabling facility - balancing the value of researcher time saved and investment in staff time at a research-enabling facility. 

\subsection*{Econometrics-based methods for estimating financial impacts and outcomes}
Empirically derived econometric estimators can be used to estimate the direct and indirect economic effects of investments in many kinds of projects and services. The Regional Input–Output Modeling System (RIMSII) uses empirically derived multipliers to estimate the total direct and indirect economic impact and the total number of jobs created by a variety of investments \cite{USBEA2024}. These multipliers are calculated for a variety of investments, ranging from building a parking garage to investing in jobs and facilities at a university, for specific geographic regions. IMPLAN (originally short for “impact analysis for planning") offers the same approach globally \cite{IMPLAN}. Because these approaches use general categories of investment, they may underestimate the value of investment in highly unusual areas. For example, when estimating the value of investment in advanced cyberinfrastructure systems within the USA, one could potentially use a RIMSII multiplier from one of two types of “Professional, scientific, and technical services,” or one of two types of “Junior colleges, colleges, universities, and professional schools.” The impact of some specialized research-enabling facility might well be greater than the calculated average for such relatively broad categories. However, these estimation methods are well known and widely trusted. They produce figures that constitute an unimpeachable lower bound for the overall economic impact of any investment. This approach has been used to demonstrate positive impact of a cyberinfrastructure center on its home state \cite{MillerStewart2014}.

\subsection*{Accounting-based methods for estimating financial impacts and outcomes}
At some level, accounting methodologies are simpler to grasp and explain than econometric methods. Fundamentally, accounting boils down to counting things and assigning values or costs to those things. There are no underlying economic theories on which accounting-based methods depend. So long as the counts are accurate and the costs and values reasonable, the results are bound to be reasonable.

\subsubsection*{Leverage}
One of the earliest concepts in accounting for the value of a particular instance of publicly funded research-enabling facility has been referred to as “leverage.” For publicly funded research facilities, leverage is simply the financial value of the sum of grant awards received by the publicly funded research projects that make use of that facility. This total value can be useful in concise statements that explain the value of a facility. Using data from \cite{Snapp-Childs2024} about how much the USA federal government invested in XSEDE and how much it invested in other grant awards that used XSEDE resources, for example, one could say that “The NSF leveraged USD 200M of investment in XSEDE to support USD 4.5B in total grant awards that made use of XSEDE facilities and services.” Indeed, this sort of statement has often been used by coauthor Towns in non-published materials prepared for the NSF as part of XSEDE project reviews.  However, these figures must be used with a measure of care because of the matter of shared credit. Suppose that a discovery critically depends on two different research-enabling facilities. The value of such a discovery can be considered as “leverage” for each of those facilities. But it would be double counting to count the full value as part of the “return on the investment" in each facility, something that has been done in some reports. It is certainly unlikely in general to be wholly true that research conducted with the use of a particular facility could or would not have been successfully carried out without the use of that facility. Such a claim might be true in the case of the research conducted with the Large Hadron Collider that led to the verification of the existence of the Higgs Boson. But, in general, one is well advised to be wary of contrapositive statements of the form “without research-enabling facility Z, discovery Beta would not have been possible.” Certainly in the case of many kinds of research-enabling facilities, some of the research done using that facility might well have been successfully carried out by clever scientists using some other facility. Some other research might have occurred, but not as efficiently or as well. And yet other research may indeed not have happened at all. So, while a “leverage” total can be useful shorthand, it is necessary to be careful when using this term. 

\subsubsection*{A proxy for return on investment (ROI)}
ROI is a cost accounting term, defined as shown in Eq 1 \cite{Kinney2011}: 

\medskip
\begin{equation}
\text{ROI}=\dfrac{\text{Income}}{\text{Assets invested}}
\end{equation}
\medskip

This equation cannot typically be applied to academic cyberinfrastructure facilities because, typically, no products or services are actually bought or sold. A \textit{proxy} for ROI has been defined \cite{Stewartetal2023} as shown in Eq 2:

\medskip
\begin{equation}
    \text{{\texorpdfstring{ROI\textsubscript{proxy}}{ROI proxy}}}=\dfrac{\text{Market value of services delivered and products created}}{\text{Cost to deliver services and products}}
\end{equation}
\medskip

This formula is essentially a way to calculate the financial efficiency of providing a service. In terms of public sector research and university activities, this formula shows whether something is being done at greater or lesser cost than would be experienced if products or services were simply bought on the open market. ROI\textsubscript{proxy} is a way to measure the financial effectiveness resulting from investments in cyberinfrastructure over any time period, in the absence of anything being bought or sold. In this formula, as in the formula for ROI, a value of 1.0 indicates a break-even level. A value greater than 1.0 indicates that the value of services provided when calculated based on market costs exceeds the actual cost of operating the facility. These calculations are conceptually straightforward. The denominator is the TCO for the relevant period of time. For the numerator, one tallies as many services and products as possible that were created using the cyberinfrastructure resources, finds the prevailing market prices for as many of these services as possible, and arrives at a total estimated market value for the relevant time period. Credible valuations are available for a surprising array of entities, including researcher time, scientific accomplishments such as a peer-reviewed publication or Nobel Prize, educational outcomes, and even improvements in quality of life. These valuations may be done annually or over longer time periods. The results are approximate, of course. Calculated results are likely to be an underestimation of the actual long-term value. There will often be benefits that are impossible to quantify and thus are not included in the calculations. Further, the economic value of some benefits will increase over time as the scope of application expands. Examples include health care and other benefits to people's quality of life. 

Much of the consideration of on-premises computing facilities has been driven by the possibility of purchasing computing services from commercial cloud providers making it useful in this paper to flesh out at least the skeleton of the calculations for ROI\textsubscript{proxy} for on-premises computing facilities vs. the use of cloud resources. The following discussion is based on prior work in \cite{Stewartetal2018}. When asking the question, “Which is more financially efficient - on-premises or cloud?” it is critical to know the total cost of ownership (TCO) for each of those two different options. 

(A quick but important note here: Although the term \textit{on-premises} is inadequate to modern data centers, it is widely used nonetheless. What this term refers to today is cyberinfrastructure possessed through ownership or long-term lease, and operated either on the premises of an organization or at a remote location either owned or leased by that organization.) 

Eq 3 shows the calculations needed to determine the cost of on-premises facilities:

\medskip
\begin{equation}
    \text{TCO\textsubscript{onprem}}=\text{AC + MC + TS + TPC + TCC}
\end{equation}
\medskip

where AC is acquisition cost, MC is maintenance cost, TS is total system administration salary, TPC is total power cost, and TCC is total co-location or space cost for the relevant time period.

TS is the sum of salary and benefits costs for FTEs (full-time equivalents) devoted to system administration for the relevant time period. 

Total power cost (TPC) as shown in Eq 4 is the product of the average consumed power P\textsubscript{avg}, the average power usage efficiency PUE\textsubscript{avg}, and the average cost per kilowatt hour kWh\textsubscript{avg}, multiplied by n, the length of the relevant time period. 

\medskip
\begin{equation}
    \text{TPC} = \text{P\textsubscript{avg}} \times \text{PUE\textsubscript{avg}} \times \text{kWh\textsubscript{avg}} \times  \text{n}
\end{equation}
\medskip

The cost for space can be amortized from the cost of a building owned by a university or calculated from co-location costs. 

The cost for cloud services can be challenging to calculate because there are so many options and pricing levels today. The pricing one chooses to use here will vary wildly depending on how the relevant agreement would be structured. If one is considering movement of workloads such as those typically run on an HPC cluster to the cloud, then the availability of cloud resources must be similar to that of a cluster resource. This means that the prices for on-demand or spot pricing are not appropriate for such an analysis. Spot pricing may work for research groups that have no deadlines, that have only short-running jobs, or that have extensive experience in checkpointing their computer programs (stopping a program in mid-execution and restarting it again later). But the job characteristics associated with use of spot pricing do not reflect the sort of resource that most researchers require in order to conduct their work effectively. Eq 5 shows the calculations - simple in and of themselves, once one has calculated the appropriate instance type and cost - that are required in order to calculate the total cost of outsourcing to a cloud provider:

\medskip
\begin{equation}
\text{TCO\textsubscript{cloud}} = \text{CH} \times \text{IC} \times \text{n}
\end{equation}
\medskip

where CH is the number of compute hours needed per relevant period of time, IC is the instance cost per unit of time, and n is the number of time units.

ROI\textsubscript{proxy} is then simply given by Eq 6:

\medskip
\begin{equation}
\text{ROI\textsubscript{proxy}} = \dfrac{\text{TCO\textsubscript{onprem}}}{ \text{TCO\textsubscript{cloud}}}
\end{equation}

A value \textgreater 1 favors on-premises as the more economical choice. Note that cost-benefit analyses are very similar and in general each of these two forms of analysis may be straightforwardly restated in the terminology of the other form of analysis.


In general, most analyses so far have shown that in terms of financial efficiency, the cost of outsourcing would be greater than the cost of on-premises facilities. This same approach with {ROI\textsubscript{proxy}} has now been applied to on-premises HPC clusters, on-premises cloud facilities, and comprehensive advanced cyberinfrastructure service delivery \cite{Stewartetal2018}. The results of these studies are discussed in more detail below.

\subsubsection*{A comprehensive framework for financial value creation}
ROI\textsubscript{proxy} is helpful in determining which financial option is most effective in operating a particular instance of a research-enabling facility. This  provides a concrete answer to the budgetary question: “Does this facility save costs relative to purchasing services at market prices?” While important when considering bottom line operational costs, the analysis cannot answer the more fundamental question: “Are the things done by this facility actually worth doing at all?”

The International Integrated Reporting Framework is of use in answering this more difficult and far more important question. The purpose of the International $<$IR$>$ Framework (as it is typically referred to) is “to explain to providers of financial capital how an organization creates, preserves, or erodes value over time” \cite{IFRSFoundation2022}. This framework includes six categories, intended to capture all possible types of value creation: financial capital, manufactured capital, intellectual capital, human capital, social and relationship capital, and natural capital. This structure, according to the IFRS (International Financial Reporting Standards) website, benefits all stakeholders, including “legislators, regulators, and policymakers.” The IFRS has a grand goal of transforming corporate reporting, so the description of their reporting framework extends well beyond what is needed for an assessment of the value of an instance of a research-enabling facility. However, the comprehensiveness of the framework, its clarity, and its backing and promotion by an international consortium of accountancy researchers and experts provide great credibility to the portion relevant to the evaluation of research facilities and research projects generally (pp. 18-23 in \cite{IFRSFoundation2022}). 

Snapp-Childs et al. \cite{Snapp-Childs2024} assessed the net value of the USA federal government investment in the large-scale XSEDE cyberinfrastructure facility to USA citizens and the USA as a whole using the International $<$IR$>$ Framework. As mentioned earlier, the clarity in accounting-based methods comes from their basic simplicity: count things, assign values, and add up the total value (or total cost). It is indeed possible to arrive at valuations for a very large range of outputs that are derived from credible sources and seem reasonable. This requires a willingness to view financial assessments as a process within which one wishes to enable the best practically possible assessments to inform decision making, rather than an insistence on the same sort of precision expected in analytical chemistry or particle physics. Snapp-Childs et al. took the approach of working with the best available estimates of value for some metrics, where reasonable assessments from seemingly credible sources were available, and simply not counting products for which assessments that seemed credible were not available. They found average financial values for a variety of entities ranging from a peer-reviewed scientific publication to a human life. They also presented a mathematical derivation for what it means for some entity to merit a percentage of the credit for an outcome brought about through a series of steps that include the involvement of that entity. Such entities range from the insight of the principal investigator to every research-enabling facility used along the way in the creation of a product. It is not possible to solve this formula analytically. However, the existence of this formula demonstrates and mathematically defines a concept that allows a statement such as the following to make sense: “XSEDE merits n\% credit for the aggregate value of the end products created as a result of the use of XSEDE services.” 

After calculating the total end-product value, Snapp-Childs et al. calculated the percentage of credit that would have to be attributable to XSEDE in order for the ROI to be at least 1.0. Starting from the premise that XSEDE deserved at least 1$\%$ of the credit for outcomes stemming from the use of XSEDE, they were able to conclude that the return on USA federal investment in XSEDE was at least 1.0. That is, XSEDE created sufficient benefits to the taxpayers of the USA and to society generally that the value created by XSEDE within the USA was greater than or equal to its cost to the taxpayers of the USA. This sort of summing up of total value and arguing based on the reasonableness of assigning a percentage of credit is not precise, but it can be persuasive when it is impossible to assign an exact percentage of the credit for various contributions to outcomes, which will generally be the case. The accounting approach set up within the International $<$IR$>$ Framework, combined with the “appeal to reasonableness” used in \cite{Snapp-Childs2024}, appears to be a reasonable and practicable option for facilities of too modest a scale to create region-wide economic impacts and/or where the economic data required to employ econometric modeling approaches is not available. Note that with this approach, economic terms such as “spillover” simply disappear from the calculations, because the $<$IR$>$ Framework is designed to capture such spillover effects directly within one of its six main categories.

\subsection*{A production-function analysis of academic and financial outputs}
Within academia, it can be useful to quantitatively examine important outputs even if it is not immediately possible to associate a financial value with them. And there are indeed, non-financial metrics within academia that are important in an institution's reputation, ranking, and its recruitment of students, faculty, and staff. In a 1928 work, Cobb and Douglas proposed a “production function” that expressed factory output (Y) as a function of investment in capital (K) and labor (L) \cite{CobbandDouglas1928}. Smith built on this approach to relate the production of a variety of academic outputs (Y) to capital investment in systems (as measured in TeraFLOPS) and labor costs for the staff that operate and support them as\cite{smith2021research}:

\medskip
\begin{equation}
\text{Y}=f{(\text{K\textsubscript{flops}}, \text{L\textsubscript{staff}}) } 
\end{equation}
\medskip

\begin{itemize}
    \item where the outputs and metrics (Y) are (in any given time period, typically annually):
    \begin{itemize}
        \item total Higher Education Research and Development (HERD) expenditures as reported to the NSF \cite{NSF2023}
        \item the total financial value of new grant awards received;
        \item the total number of peer-reviewed publications;
        \item the total number of high-impact publications (as listed in \cite{SpringerNature2025});
        \item the total number of earned doctorates, as reported to the NSF National Center for Science and Engineering Statistics \cite{NCSES2023}
\end{itemize}
\item and the inputs are:
    \begin{itemize}
        \item K\textsubscript{flops}: total TeraFLOPS operated on premise (that is, total processing capability of local HPC systems in trillions of floating point operations per second, a time-dependent proxy for capital);
        \item {and L\textsubscript{staff}: salary costs(labor) for RCD professionals}.
    \end{itemize}
\end{itemize}

All of the outputs are in some measure a function of investment in cyberinfrastructure. HERD expenditures and new grant awards are related to each other over time but, within any given year, relatively distinct. HERD expenditures in a given year represent total research and development expenditures, most of which are a result of grant funding in prior years but some of which may also reflect expenditures from gifts other than grants as well as the use of institutional funds. New grant award dollar totals in a given year are a measure of success in obtaining funding in the form of extramural grants specifically. The vast majority of such funds will be spent in the years following the initial award year. Conceptually, HERD expenditures are a function of investment in cyberinfrastructure, because the ability to execute externally funded research in many sciences (and some humanities and arts) is a function of the local computing power required to perform analyses, do simulations, and create visualizations. The availability of local cyberinfrastructure is often a consideration in the evaluation of grant proposals in these same fields, so success in grant-getting can depend in part on local cyberinfrastructure resources.

Smith’s 2024 analysis considered over 21 years of data from Purdue University. Key results from this analysis are shown in Tables 1 and 2.

\begin{table}[!ht]
\begin{adjustwidth}{-.5in}{0in} 
\centering
\caption{
{\bf Relationship of academic output metrics to on-premises HPC TeraFLOPS capacity and investment in staff supporting use of HPC systems at Purdue University, 1999–2020. }}
\begin{tabular}{|l+l|l|l|l|l|l|l|}
\hline
& \multicolumn{5}{|l|}{\bf Fitted Slope of Outputs as Linear Correlates}\\ 
& \multicolumn{5}{|l|}{\bf of Inputs, Measured Annually}\\ 
\hline
&  {\bf HERD} &{\bf New}  &{\bf Publications} & {\bf High-} & {\bf PhDs}\\ 
&  {\bf Expenditures} &{\bf Grant}  & & {\bf Impact} & {\bf Awarded}\\ 
&  {\bf in USD Ms} &{\bf Awards} & & {\bf Publications}  & \\
&  {\bf } &{\bf in USD Ms}  & &  & \\
\thickhline
$\textit{100s}$ $of$ & 1.29* & 1.81* & 7.14 & 3.34** & 1.84** \\ 
$TeraFLOPS$ & & & & &  \\ 
\hline
\textit{USD 100Ks} $of$ & 14.34*** & 9.56** & 155.30*** & 22.98*** & 10.23*** \\ 
$Salaries$ $of$&  &  &  &  &   \\ 
$RCD$ &  &  &  &  &   \\
$Professionals$ &  &  &  &  &   \\
\hline

\end{tabular}

\begin{flushleft} Table notes: * indicates statistical significance of correlation at the p$<$0.05 level; ** indicates p$<$0.01 level; and *** indicates p$<$0.001 level. Research Computing and Data (RCD) professional staff investment is listed as a function of USD 100Ks in salary because this is roughly equivalent to one full-time staff person.
\end{flushleft}
\label{table1}
\end{adjustwidth}
\end{table}

\begin{table}[!ht]
\begin{adjustwidth}{-.5in}{0in} 
\centering
\caption{
{\bf Percent variance in academic output metrics at Purdue University, explained by on-premises HPC TeraFLOPS capacity and investment in staff supporting use of HPC systems at Purdue University, 1999–2020. }}
\begin{tabular}{|l+l|l|l|l|l|l|l|}
\hline
& \multicolumn{5}{|l|}{\bf Fitted Slope of Outputs as Linear Correlates}\\ 
& \multicolumn{5}{|l|}{\bf  of Inputs, Measured Annually}\\ 
\hline
&  {\bf HERD} &{\bf New}  &{\bf Publications} & {\bf High-} & {\bf PhDs}\\ 
&  {\bf Expenditures} &{\bf Grant}  & & {\bf Impact} & {\bf Awarded} \\ 
&  {\bf in USD Ms} &{\bf Awards} & & {\bf Publications}  & \\
&  {\bf } &{\bf in USD Ms}  & &  & \\
\thickhline
$\textit{100s}$ $of$ & 31\%  & 34\% & 29\% & 35\% & 36\% \\ 
$TeraFLOPS$ & & & & &  \\ 
\hline
\textit{USD 100Ks} $of$ & 63\% & 54\%  & 65\% & 61 \% &  58\%\\ 
$Salaries$ $of$&  &  &  &  &   \\ 
$RCD$ &  &  &  &  &   \\
$Professionals$ &  &  &  &  &   \\
\hline
$Other$ & 6\% & 12\% & 5\% & 4\% & 6\%  \\ 
$Factors$&  &  &  &  &   \\ 

\hline
\end{tabular}

\begin{flushleft} Table notes: Research Computing and Data (RCD) professional staff investment is listed as a function of USD 100Ks in salary because this is roughly equivalent to one full-time staff person.
\end{flushleft}
\label{table2}
\end{adjustwidth}
\end{table}

The strength of the relationship between investment in cyberinfrastructure hardware and professional RCD staff is impressive - almost surprising. Relevant to this is the fact that during the time periods measured all of the other major research-enabling facilities (e.g. farming facilities for the School of Agriculture) stayed relatively consistent over time. This analysis has been extended across five different R1 institutions \cite{Smithetal2025}. As shown in Tables 3 and 4, the correlations are somewhat weaker across five institutions than within one. Still, over a total of 82 total years of data from five institutions, strong relations are detectable.

The sorts of production-function analyses shown here are not directly translatable into financial terms that extend beyond academia in their impact. However, within academia, the strong relationships shown between investment in cyberinfrastructure facilities coupled with support services and important academic metrics is likely to impress academic deans in the sciences, vice presidents and vice chancellors for research, presidents, and possibly even CFOs. The financial relationships should be impactful for most stakeholders within academia and many in government.

\begin{table}[!ht]
\centering
\caption{
{\bf Relationship of academic output metrics to on-premises HPC TeraFLOPS capacity and investment in staff supporting use of HPC systems at 5 R1 universities. }}
\begin{tabular}{|l+l|l|l|l|l|l|l|}
\hline
& \multicolumn{4}{|l|}{\bf Fitted Slope of Outputs as Linear Correlates}\\ 
& \multicolumn{4}{|l|}{\bf  of Inputs, Measured Annually}\\ 
\hline
&  {\bf HERD} &{\bf Publications} & {\bf High-} & {\bf PhDs}\\ 
&  {\bf Expenditures}  & & {\bf Impact} & {\bf Awarded}\\ 
&  {\bf in USD Ms} & & {\bf Publications}  & \\

\thickhline
$\textit{100s}$ $of$ & 3.10** & 24.63* & 4.47** & 1.56 \\ 
$TeraFLOPS$ & & & &  \\ 
\hline
\textit{USD 100Ks} $of$ & 14.46*** & 134.97*** & 20.78*** & 7.68*** \\ 
$Salaries$ $of$&  &  &  &    \\ 
$RCD$ &  &  &  &    \\
$Professionals$ &  &  &  &   \\
\hline

\end{tabular}

\begin{flushleft} Table notes: * indicates statistical significance of correlation at the p$<$0.05 level; ** indicates p$<$0.01 level; and *** indicates p$<$0.001 level. Research Computing and Data (RCD) professional staff investment is listed as a function of USD 100Ks in salary because this is roughly equivalent to one full-time staff person.
\end{flushleft}
\label{table4}
\end{table}


\begin{table}[!ht]
\centering
\caption{
{\bf Percent variance in academic output metrics explained by on-premises HPC TeraFLOPS capacity and investment in staff supporting use of HPC systems at 5 R1 universities. }}
\begin{tabular}{|l+l|l|l|l|l|l|l|}
\hline
& \multicolumn{4}{|l|}{\bf Fitted Slope of Outputs as Linear Correlates}\\ 
& \multicolumn{4}{|l|}{\bf of Inputs, Measured Annually}\\ 
\hline
&  {\bf HERD}  &{\bf Publications} & {\bf High-} & {\bf PhDs}\\ 
&  {\bf Expenditures}  & & {\bf Impact} & {\bf Awarded} \\ 
&  {\bf in USD Ms}  & & {\bf Publications}  & \\
&  {\bf }   & &  & \\
\thickhline
$\textit{100s}$ $of$ & 18\%  & 15\% & 17\% & 1\% \\ 
$TeraFLOPS$ & & & & \\ 
\hline
\textit{USD 100Ks} $of$ & 44\% & 39\%  & 40\% & 25\%  \\ 
$Salaries$ $of$&  &  &  &    \\ 
$RCD$ &  &  &  &    \\
$Professionals$ &  &  &  &   \\
\hline
$Other$ & 38\% & 46\% & 43\% & 74\% \\ 
$Factors$&  &  &  &   \\ 

\hline
\end{tabular}

\begin{flushleft} Table notes: Research Computing and Data (RCD) professional staff investment is listed as a function of USD 100Ks in salary because this is roughly equivalent to one full-time staff person.
\end{flushleft}
\label{table3}

\end{table}

\section*{Discussion}
This paper has a twofold purpose: to demonstrate the effectiveness of several approaches, many new, to assessing the value of investment in cyberinfrastructure with a focus on financial measurements; and to discuss the extension of these methods to other kinds of research-enabling facilities and research projects more generally.

\subsection*{Evaluation of investment in cyberinfrastructure facilities}
The core, long-standing, and common result of “lease vs. buy” analyses is that it is financially more efficient to operate on-premises research computing systems (or, as noted earlier, to house university-owned systems in owned or leased remote data centers). That is, ROI\textsubscript{proxy} for on-premises operations is well over 1. For example, \cite{Stewartetal2018} found in 2018 that the cost to outsource to a cloud provider would have been between 2.5 and 3.7 times greater than the cost of owning and operating an on-premises supercomputer. They found roughly similar results when comparing outsourcing with an on-premises cloud system - outsourcing being 1.9 to 3.1 times more expensive than an on-premises cloud. Several analyses of the cost-effectiveness of XSEDE now consistently show a calculated ROI\textsubscript{proxy} well over 1. That is, it is far less expensive for the USA federal government to offer grants and contracts to universities to provide cyberinfrastructure services than it would be to purchase such services on the open market within the USA. These ROI figures are somewhat higher than the figure of 1.8 found in an analysis of a recent upgrade to CERN facilities \cite{Florio2019}. However, considering the differences in methodology, this must be considered a reasonable consistency among these findings. 

The strongest claim about the societal value of cyberinfrastructure that has been made to date in the USA is the assertion that the total benefit of XSEDE to USA taxpayers and USA society as a whole has been greater than the cost of XSEDE to the USA federal government \cite{Snapp-Childs2024}. This argument depends, to a certain extent, on an appeal to reasonableness. Snapp-Childs et al. used the seemingly low figure of 1\% as a minimum estimate XSEDE's contribution to the total outputs created with the aid of XSEDE resources and services. The arguments presented in that paper seemed reasonable at the time of publication and still do. This analysis is, to the best of our knowledge, the first time that anyone has quantitatively evaluated the national societal and economic value of a cyberinfrastructure facility and numerically argued a net benefit to taxpayers.

The production-function analysis approach pioneered by Smith is the newest of the major methods for assessing investment in cyberinfrastructure. It has proven to be an excellent mechanism for documenting relationships between investment in cyberinfrastructure and important returns within institutions of higher education.

Looking at Purdue University specifically, the analysis presented by Smith offers clear guidance regarding the benefits of investing in equipment and staff. Today, USD 100,000 will pay for compute node servers providing approximately 200 TeraFLOPS of FP64 Tensor Core (a specific nVidia product described at \cite{nvidia2025}) performance. A leader at Purdue University could conclude that an investment of USD 100,000 in computational capability will correspond with an increase in total university research expenditures by the institution of USD 6.45M. This implies a significant increase in receipts of extramural grant funding. Similarly, USD 100,000 is roughly the full annual cost of one entry-level RCD professional (salary and benefits). Every USD 100,000 invested in staff costs – adding one person – corresponds to an increase of USD 14M in total grant expenditures. Indeed, at Purdue University, which is strongly focused on engineering, the analysis by Smith suggests that as of the time of his analysis, access to computing power was an enabler and a limiting factor of research activity at a very large scale within the university community. The fact that relationships between academic and financial output metrics and independent variables related to investment in computational facilities and RCD staff remain statistically significant when analyzed across six different R1 universities is a strong testament to the reality of an underlying causal relationship. Of course, what will be meaningful is the relationship within a particular university. For higher education leaders trying to react sensibly and strategically to budget challenges, the production-function approach seems to have much to offer. 

The results regarding ROI\textsubscript{proxy} for on-premises cluster facilities compared to cloud services consistently show that on-premises facilities are more cost effective. But this is specific to cases in which the workloads being supported run effectively on on-premises clusters. Looking at cases where a choice is being made that is not financially efficient can produce insights regarding the function of research facilities.

There are relatively few peer-reviewed analyses of the academic uses of commercial cloud services as a research-enabling facility. (Publications that give the results of surveys by cloud vendors and so-called industry observers are easy to find, but they are, typically, essentially advertisements, not scholarly research.) Three recent systematic and peer-reviewed surveys of commercial cloud use are \cite{CASCsurvey2020}, \cite{CASCSurvey2021}, and \cite{Thakur2022}. The Coalition for Academic Scientific Computation (CASC) is an organization focused on the research uses of cyberinfrastructure in higher education in the USA \cite{CASC2025}. The two CASC surveys focus on use of commercial clouds in supporting research activities. These surveys have confirmed a rise over time in the academic use of cloud facilities in support of research activities. The 2021 survey found that the reasons for such use were primarily related to functional issues rather than cost. In that survey, the most frequent reasons given for the purchase of commercial cloud resources in support of research activities were (in decreasing order of frequency, out of a total of 69 responses):
\begin{enumerate}
   \item “experimentation with utility for research”;
   \item “followed fairly distantly by a cluster of three answers:” 
    \subitem “need for cloud-native capabilities”;
    \subitem “autonomy for researchers to make
    their own choices”;
    \subitem “need for \textit{always on} capability'”;
   \item “10 institutions report making use
   of commercial cloud resources in order to save money at least in some circumstances.”
\end{enumerate}

On-premises clouds are another issue. They are more cost effective than commercial clouds, according to one study \cite{Stewartetal2018}, but operating an on-premises cloud requires a much greater investment in staff and a greater scale of local resource than do local clusters or the use of commercial clouds. When something is more cost efficient but out of the scope of an institution's needs or financial capabilities, financial efficiency is irrelevant. 

The above observations are of course specific to cyberinfrastructure, but they help make a point about the study of research facilities generally. Money \textit{isn't} everything. Understanding what research institutions and researchers are doing when they are not taking what appears to be the most financially efficient path can help provide some insight into other factors in the usage of research facilities in general.

\subsection*{Assessment of research value in the Global South and Global North}

Availability of public funding is a key enabler of publicly funded research, and as such there has historically been much less such research in the Global South (cf. \cite{Confrariaetal2016}). In fact, in some areas there has been considerable progress in the publication of and citations to research led by researchers in the Global South (for example, in terms of sustainability research, see \cite{Dangles2022}). There have been notable publications taking a return on investment approach to evaluating medical interventions in the Global South. For example, \cite{Nicholson2024} 
 analyzed the cost-effectiveness of surgical interventions in patient care in a conflict-affected area of South Sudan and found a ROI of 14. ROI for healthcare intervention in the form of deploying health extension workers in Ethiopia found an ROI of 1.3 to 3.7 \cite{Bowser2023}. We have not found any ROI analyses for research activities in the Global South, but clearly this concept is well understood and used. The quantitative methods described in this paper were a good deal of work to develop and deploy in the form of first proofs of concept. Now that this has been done, applying them will, we hope, be more straightforward. It is the hope of the authors that the methods presented here will be of use to researchers and research administrators in the Global South as they seek to justify additional funding for research about the Global South, led by researchers from the Global South. This hope is not to imply that the Global South is itself homogeneous but there are many problems faced in the Global South where researchers and countries share more in common with each other than with researchers and countries in the Global North.

As regards the Global North, there are many reasons to hope that the current state of uncertainty about funding for university-based research comes to an end. By the same token, there are many reasons to hope that the wave of higher ed closures seen in the USA is not replicated in the EU. We hope that the presentation of methods for quantitatively expressing the value of university-based research institutions - straightforwardly extensible to university-based research generally, will become useful tools in addressing current questions about the value of such research. A reasonable critic might offer the observation that some perhaps significant portion of today’s skepticism regarding higher education comes from a failure in the past to analyze cost-benefit ratios according to different perspectives and value systems. Indeed, in the USA, it was possible in the 1980s to justify the then-new NSF supercomputer program entirely based on the potential new knowledge it would create. The creation of new knowledge, not possible to create without use of a supercomputer, was considered sufficient without any reference to cost/effectiveness at all \cite{CI-Encyclopedia2018} that was impossible to obtain without use of a supercomputer. The methods demonstrated here might, if widely applied, lead to greater consensus regarding at least this aspect of higher education institutions in the Global North.

\subsection*{Extensions to other research-enabling facilities and research generally}
Decades after the initial rise to popularity of the total quality management approaches discussed earlier in this paper, Kaplan and Norton tackled the challenge of conveying information about organizational performance clearly and concisely. They developed the Balanced Scorecard in order to present “a set of measures that gives top managers a fast but comprehensive view of the business” (Kaplan and Norton 1992). The Balanced Scorecard focuses on four questions, cast here in the form of questions about a research-enabling facility:
\begin{itemize}
    \item How do customers see the facility? (customer perspective);
    \item What must the facility excel at? (internal perspective);
    \item Can the facility continue to improve and create value? (innovation and learning perspective); 
    \item How does the facility look to shareholders? (financial perspective).
\end{itemize}

Fig~\ref{fig1} presents an example of Balanced Scorecard template for the assessment of research-enabling facilities. This example is as general as possible. This depiction shows rows of metrics applicable to any sort of research-enabling facilities where general exemplar metrics can be suggested. In places where metrics must be more specific in order to make sense, for example in operational metrics, examples from cyberinfrastructure facilities are shown. Such operational metrics can be altered to express the essential measures for any kind of research-enabling facilities. Following Kaplan and Norton's discussion, all metrics should somehow be related to organizational strategies and tactics and may include statements of quantitative goals set (e.g., key performance indicators). Note that the arrows are as shown in Kaplan and Norton's original example and indicate the interrelatedness of organizational performance in the four general categories they specify.  This example, with little modification, should be useful to anyone trying to argue for the value of any cyberinfrastructure facility that supports research. 

More broadly, the Balanced Scorecard concept and structure can be used by anyone trying to justify any sort of research-enabling facility and should help them to hone their analysis and presentation. The higher education ecosystem is under pressure in the USA today in a way that is unprecedented since the end of World War II. To be deemed effective, answers about the value of research-enabling facilities presented to stakeholders must be presented within the value set of the stakeholders. The methods discussed here address many perspectives, including the perspective of someone whose only frame of reference is based on money and economic growth. For others, such as leaders in higher education who are making tough budget-allocation decisions, some of the more academically oriented metrics presented here may be of greater interest. The Balanced Scorecard concept presents a way to tie together the best parts of the forms of analysis discussed in this paper.


\begin{figure}[!h]
\caption{\bf Balanced Scorecard for research-enabling facilities. Metrics applicable to any sort of research-enabling facilities are shown in plain text. Italics are used for operational metrics which are by their nature specific to a particular sort of facility. \textsuperscript{1}It is generally possible to estimate total societal value only after a facility has been in operation for some time. \textsuperscript{2} There will be discipline-specific operational metrics relevant to the mission of any research-enabling facility.}
\label{fig1}

\end{figure}

\subsection*{Opportunities for future research}
There have been significant advances in the methodologies for assessing research facilities in recent years, largely focused, at least in the USA, on cyberinfrastructure. There are still more topics to explore and additional opportunities for further developments, including the following:
\begin{itemize}
    \item Impact on STEM workforce development. Miles et al. considered how cyberinfrastructure supports the career development of a science, technology, engineering, and mathematics (STEM) workforce \cite{Milesetal2023}. Some steps to further quantify this relationship have already been published \cite{Snapp-Childsetal2025}, but there is considerable need for additional research on this topic.
    \item The assertion mentioned earlier that there is a linkage, among otherwise similar papers, between the use of advanced computing facilities in the preparation of a paper and the long-term impact of that paper merits further investigation.
    \item The approach used by Wiedmann \cite{Wiedmann2017} to quantify the value of research-enabling facilities in terms of time saved merits further attention. 
    \item “Real Options” is a financial evaluation approach discussed in 2008 in \cite{vanRhee2008}, which has so far not been used in the evaluation of research facilities but appears to have the potential for utility in such analyses.
    \item  Newly developed frameworks for metrics regarding cyberinfrastructure centers were presented in a recent online journal topical collection \cite{Aurora2024}. These all merit further development. Further, these metrics could possibly be conveyed very effectively and concisely if presented in a Balanced Scorecard structure.
    \item Relationship of assessment and achievement in countries and areas where there is a demand to expand research. The research cited in this discussion is drawn exclusively from the USA and EU, where the challenge perceived by academics and policy makers is very often to preserve a privileged position in research competitiveness. Countries in the Global South may very well feel a desire or need to expand research capabilities.  The methods presented here, particularly those applicable on a basis of annual evaluations, seem potentially as useful in a situation in which there is a desire to expand research activities as one where there is a desire to preserve levels of investment, but to our knowledge this sort of analysis either has not been done or has rarely been done. Applicability of the techniques presented here in other contexts may be the single most important opportunity for future research.
\end{itemize}

As long as finances remain a critical concern in higher education - and there is no end in sight to concerns about finances in higher education - financially based assessments are likely to be of value in practical decision making and in conveying the value of investments in research facilities and research generally to the taxpaying public. It seems highly likely that it will be valuable to continue extending the methods of evaluation.

\section*{Conclusions}
Credible hypotheses about the linkage between economic growth and investment in innovation have existed at least since Schumpeter’s 1912 \textit{Theorie der wirtschaftlichen Entwicklung} \textit{(Theory of Economic Development)}. Recent analyses qualitatively suggest a relationship between investment in information technology and economic benefits at a national level \cite{NASEM2020}. It is far more difficult to assess the value of any particular instance of a research-enabling facility, such as a cyberinfrastructure facility. However, doing so is essential for setting policy and investment plans for any type of infrastructure at any given institution or within any given national research sector. Even more essential today may be using such tools to demonstrate the financial value of such investments to potentially skeptical stakeholders, including elected officials and taxpayers.

We have presented here several approaches to assessing and conveying the value of investment in facilities. 
We can summarize these approaches as follows:
\begin{itemize}
    \item Long-established methods for total quality management analyses, which are essential non-financial metrics of facility performance;
    \item Bibliometric approaches, measuring intellectual impact within scientific and technical communities;
    \item Assessment of time savings;
    \item Econometrics-based methods for estimating financial impacts and outcomes, which can provide an unimpeachable lower bound for the total regional economic impact of a research-enabling facility;
    \item Accounting-based methods for estimating financial impacts and outcomes:
     \begin{itemize} 
         \item Leverage, useful for shorthand statements but to be used with care;
         \item A proxy for ROI, well-suited to assessing the cost-effectiveness of research-enabling facility operations and which can be measured annually;
        \item The International $<$IR$>$ Framework, which addresses the total societal value of investments in a research-enabling facility but is only straightforward to implement, most likely, for long-term projects;
    \end{itemize}
    \item Production-function analyses, which can demonstrate relationships between investments in facilities and academic and financial outputs. Such analyses can be uniquely valuable for policy setting and management within organizations.
\end{itemize}

These methods may be applied to any type of research-enabling or research-facilitating infrastructure. Some of these methods are of interest only within academia but stand to be highly valuable, within those academic institutions, for decision making about research-enabling facility investments. Many of the above methods address the financial value of investment in research facilities at the regional or national level. These methods may be valuable when attempting to convey, to stakeholders ranging from individual voters to elected officials, the value - to society, to the economy, or to the global competitiveness of a nation as a whole - of investment in research. Most of these methods may be applied retroactively, given the existence of sufficiently good recordkeeping. These approaches, combined with the Balanced Scorecard as a way to depict outcomes, offer important tools for research policy planning and analysis. These tools also offer the potential to make convincing a case for the importance of public investment in research and research-enabling facilities even to constituents and stakeholders who will consider only returns stated in financial terms as significant. As institutions of higher education in the USA and EU face important and difficult decisions regarding allocation of funds internally, and skeptical stakeholders externally, clarity in demonstrating value is important. In the Global South, research institutions and institutions of higher education dedicated to research programs that further advances in quality of life and in the economy may find these techniques applicable in arguing for expanded investment in research. Stakeholder support for government funding of research depends on many factors, of which one essential factor is clarity in expressing the value of research carried out by institutions of higher education.

\paragraph*{S1 Table 1.}
\label{S1_Table}
{\bf Universities in the USA announcing annual budget deficits of more than USD 10M in 2023–2024.}






\section*{Acknowledgments}
Thanks to Kristol Hancock, Marie Deer, and Tracey Metivier for editing. Thanks to Katja Bookwalter, who created Figure 1. The research presented here was supported in part by grant awards from the USA National Science Foundation (awards \#1548562, \#2227627, \#1445604, \#2005506, \#2005632) and has also been supported in part by the Indiana University Pervasive Technology Institute. Any opinions expressed are those of the authors and may not necessarily represent the views of any funding organizations.

\section*{Declarations}
Data availability statement: No new data presented. Data sources for works cited by coauthors are available from sources cited in those papers/  Author contributions (CRediT): Conceptualization - all; Formal analysis - all; Funding acquisition - Snapp-Childs, Smith, Towns; Investigation - all; Writing original draft, review, editing - all. Generative AI statement: No generative AI used. 

\nolinenumbers

%
%
%

\bibliography{bibliography.bib}




\end{document}